\documentclass{PoS}

\title{The width of the $\omega$ meson in dense matter \footnote{
This work is partly supported by FIS2011-28853-C02-01, FIS2011-24154 and FPA2010-16963, by the Generalitat
Valenciana under Prometeo grant 2009/090, by
Grant No. 2009SGR-1289 from Generalitat de Catalunya, from FP7-PEOPLE-2011-CIG under
contract PCIG09-GA-2011-291679, and HadronPhysics3 Grant Agreement
n. 283286 under the EU FP7 Programme.
}}

\ShortTitle{The width of the $\omega$ meson in dense matter}

\author{\speaker{Laura Tolos}\\
       Instituto de Ciencias del Espacio (IEEC/CSIC) Campus Universitat Aut\`onoma
de Barcelona, Facultat de Ci\`encies, Torre C5, E-08193 Bellaterra (Barcelona), 
Spain\\
Frankfurt Institute for Advanced Studies (FIAS). Johann Wolfgang Goethe University. Ruth-Moufang-Str. 1. 60438 Frankfurt am Main. Germany\\
        E-mail: \email{tolos@ice.csic.es}}
\author{Raquel Molina\\
    Research Center for Nuclear Physics (RCNP),
Mihogaoka 10-1, Ibaraki 567-0047, Japan  } 
        \author{Eulogio Oset\\
   Departamento de F\'{\i}sica Te\'orica and IFIC, Centro Mixto Universidad de 
Valencia-CSIC,
Institutos de Investigaci\'on de Paterna, Aptdo. 22085, E-46071 Valencia,
Spain}     
        \author{Angels Ramos\\
       Departament d'Estructura i Constituents de la Mat\`eria and Institut de
Ci\`{e}ncies del Cosmos, Universitat de Barcelona, Avda. Diagonal 645, E-08028
Barcelona, Spain }

\abstract{We obtain the width of the $\omega$ meson in dense nuclear matter by taking into account (i)
 the free decay of the $\omega$ into three pions, which
is dominated by $\rho \pi$ mode, (ii) the processes induced by a vector-baryon
interaction dominated by vector meson exchange, and (iii) the $\omega \to K \bar K$ mechanism in matter. The $\omega$ meson develops an important width in matter,  coming from the dominant
$\omega~\to~\rho\pi$ decay mode, with a value of $121 \pm 10$
MeV at normal nuclear matter density for an $\omega$ at rest. At finite momentum, the width of the $\omega$ meson increases moderately with values of 200~MeV at 600~MeV/c.
}

\FullConference{XV International Conference on Hadron Spectroscopy-Hadron 2013\\
		4-8 November 2013\\
		Nara, Japan }

\begin{document}

\section{Introduction}
\label{Intro} 

The interaction of vector mesons with nuclei has been a matter of much attention over the past decades. One of the more thoroughly investigated vector mesons is the $\omega$ meson. 

From the experimental point of view, there are several investigations on the properties of the $\omega$ meson in matter with proton beams on nuclei at KEK by E325 Collaboration \cite{Ozawa:2000iw}, photoproduction on nuclei by CBELSA/TAPS \cite{Trnka:2005ey}, photonuclear reactions looking for dileptons in the final state by CLAS \cite{Wood:2010ei} or dilepton production in p+p and p+Nb at HADES \cite{Agakishiev:2012vj}. These experiments seem to point to the existence of a large width of the $\omega$ meson in the medium.

Different scenarios are present in the theoretical determination of the $\omega$ properties in matter. The obtained mass shifts range from an attraction of the order of 100-200 MeV
\cite{Klingl:1998zj,tsushima}, through no changes in the mass
\cite{Muhlich:2003tj}, to a net repulsion \cite{Lutz:2001mi}.  As for the in-medium width of an $\omega$ meson at rest, the models of \cite{Klingl:1998zj,Post:2000rf} reported a value of about 40~MeV, while the width was found to be around 60~MeV in \cite{Riek:2004kx}. All these studies show a considerable increase of the $\omega$ width in the medium.

In this paper we study the $\omega$ width in dense matter, similary to the $\bar K^*$ meson \cite{Tolos:2010fq,review}, paying a special attention to the decay of the $\omega$ into three pions via the dominant $\rho \pi$ decay mode \cite{omega}.

\section{Formalism: The $\omega$ self-energy in matter}
\label{form}

\begin{figure}[h]
\begin{center}
\includegraphics[width=0.5\textwidth,height=0.11\textwidth]{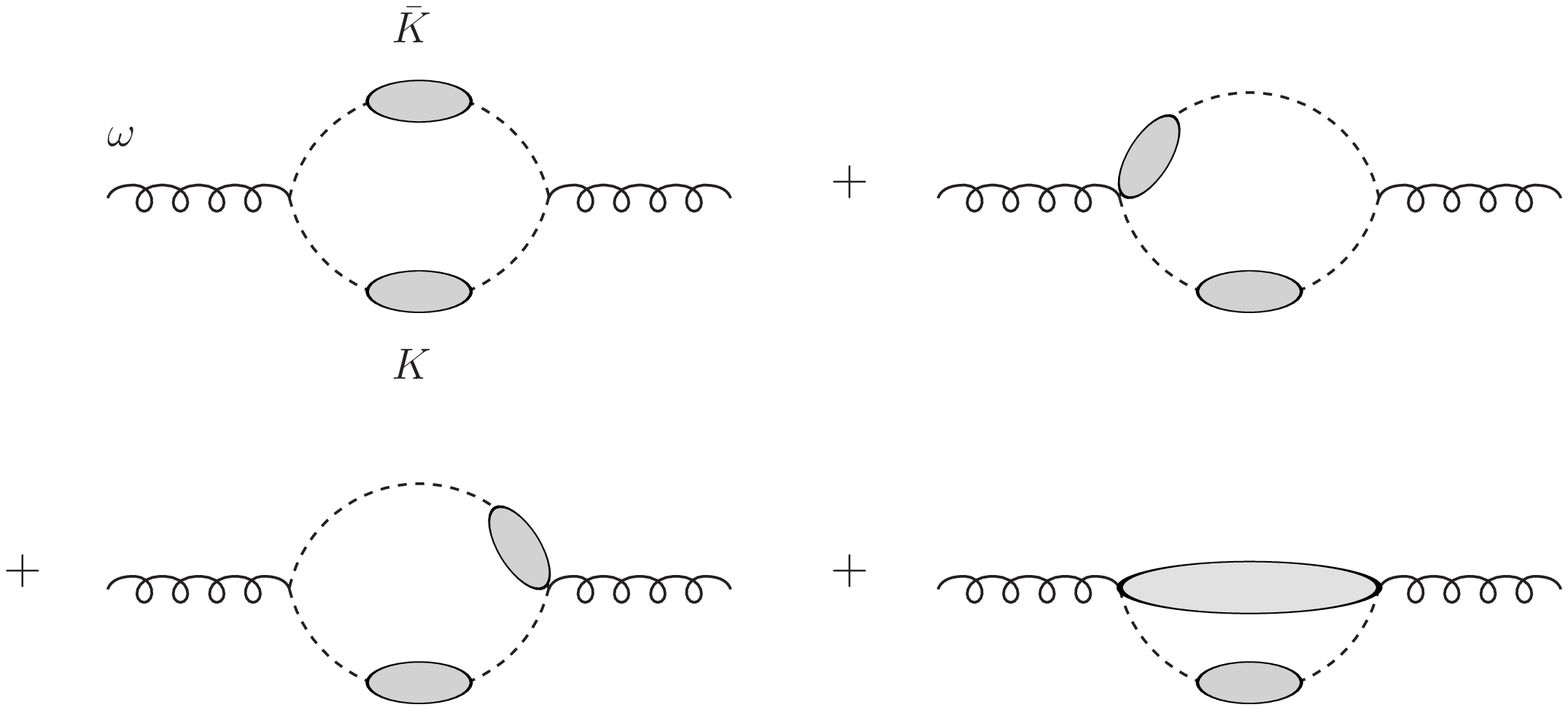}
\includegraphics[width=0.3\textwidth,height=0.12\textwidth]{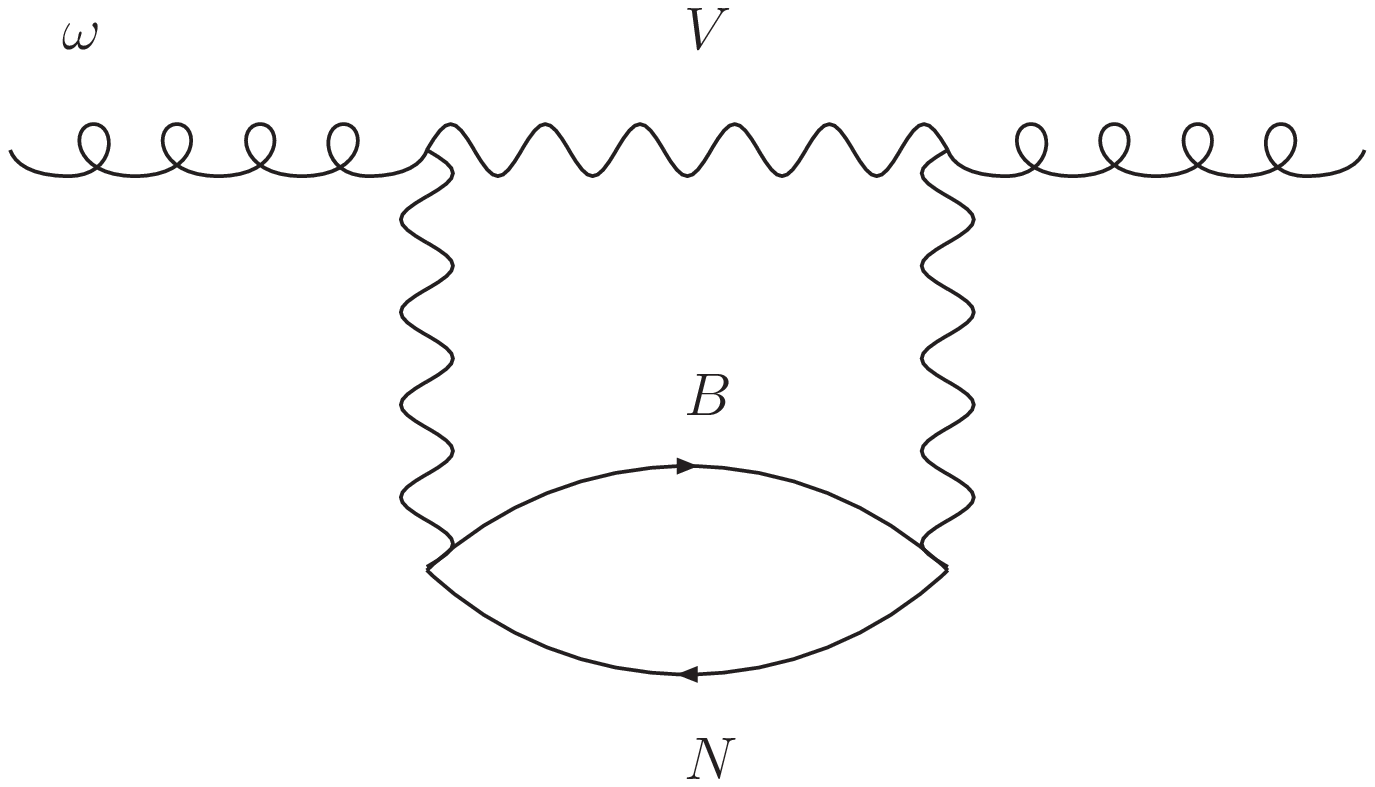}
\caption{The $\omega$ self-energy from the $\omega\to \bar{K}K$ channel in the nuclear medium including vertex corrections  for the antikaon (left plot),  and from the s-wave $\omega N$ interaction with
vector mesons and baryons (right plot). }
\label{fig:diag}
\end{center}
\end{figure}

\begin{figure}[h]
\includegraphics[width=0.8\textwidth,height=0.2\textwidth]{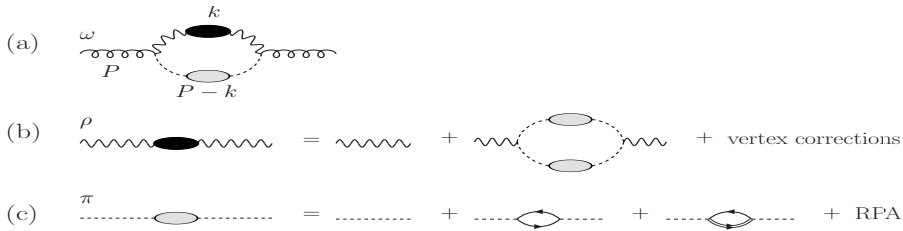}
\caption{The $\omega$ meson self-energy 
from its decay into the $\rho\pi$ (a), where the $\rho$ meson decays into
two pions (b) and the $\pi$  is dressed by its coupling
to particle-hole and $\Delta$-hole  including short-range
correlations (c).}%
\label{fig:self_rhopi}%
\end{figure}

A free $\omega$ meson decays predominantly into three pions, most of the strength
associated to the $\omega \to \rho \pi$ process with the subsequent decay of the $\rho$ meson into two pions. The
$\omega$ width is small, $\Gamma^{(0)}_\omega=8.49\pm0.08$ MeV, with 89.2\% of this value corresponding to the $3\pi$ decay channel.  This is due to the fact that the $\omega \to \rho \pi$ mechanism proceeds through the tail of the $\rho$-meson distribution. The situation, however, changes drastically in the nuclear medium. 

First, the $\omega\to K \bar K$ mechanism is energetically open in matter when the medium modifications of the $\bar K$ and $K$ mesons are incorporated (see left plot of Fig.~\ref{fig:diag}). The $\bar{K}$ self-energy in matter is obtained from the $\bar K N$ interaction within a chiral unitary approach ~\cite{angels,Tolos:2006ny,Tolos:2008di}. For $K$, due to the much weaker $KN$ interaction, we use the low-density approximation \cite{Oset:2000eg,Tolos:2003qj}. Moreover, because of gauge invariance of the model, it is necessary to include vertex corrections.

Second,  the $\omega$ properties are modified due to quasielastic and inelastic vector-baryon processes dominated by vector meson exchange.  The contribution to the $\omega$ self-energy
coming from the s-wave $\omega N$ interaction with vector mesons and baryons is depicted on the right plot of Fig.~\ref{fig:diag}. The $\omega N$ interaction is constructed within the hidden gauge formalism
in coupled channels \cite{Oset:2009vf}.  The vector meson-baryon scattering amplitudes are then obtained from the coupled-channel on-shell Bethe-Salpeter equation by incorporating medium modifications on the intermediate  states \cite{omega}.

Finally, the most important contribution to the $\omega$ width in matter comes from its
decay into  $\rho\pi$ in the nuclear medium due to the increase of the phase space available as compared to the free case. 
The self-energy for the $\omega \to \rho \pi$ process is depicted in Fig.~\ref{fig:self_rhopi}(a), where
the $\rho$- and $\pi$-meson lines correspond to their medium propagators shown in Figs.~\ref{fig:self_rhopi}(b) and (c), respectively. The pion in matter  is dressed via its self-energy which is strongly dominated by the $p$-wave coupling to particle-hole and $\Delta$-hole components and also contains  a small
repulsive $s$-wave contribution, as well as  short-range correlations and contributions from 2$p$-2$h$ excitations.
For the $\rho$-meson we employ three different  self-energy models, as we will see.

Note that in our calculation in matter we do not consider interference terms between the different physical states $\rho ^+ \pi^-$, $\rho ^+ \pi^-$ and $\rho^0 \pi^0$. While in free space, we miss an important part of the free $\omega$ width, the interference terms are negligible in matter \cite{omega}. Moreover, we also need to incorporate the contribution of uncorrelated three pions. This contribution can be supplied by either introducing a contact term that provides a background  to be added to the $\omega \to \rho \pi$ process, as done in Ref.~\cite{Gudino:2011ri}, or by adjusting the coupling of  $\omega \to \rho \pi$  to reproduce the complete free $\omega \to \pi\pi\pi$ width directly from the $\rho \pi$ mechanism.  We analyze both mechanisms in the following.

\section{Results: The width of the $\omega$ meson in  matter}
\label{res}

\begin{figure}[t]
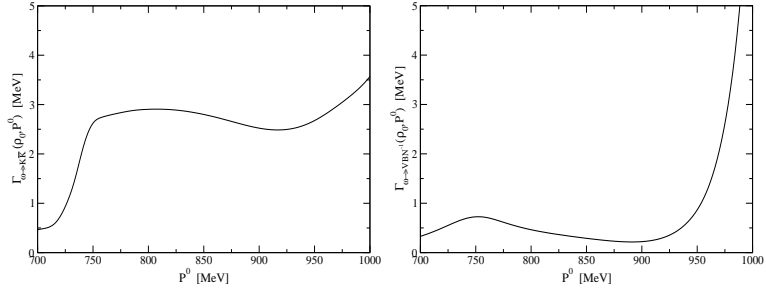

\begin{center}
\includegraphics[width=0.33\textwidth,height=0.25\textwidth]{fig12_omekkbar.eps}
\includegraphics[width=0.33\textwidth,height=0.25\textwidth]{fig13_omevb.eps}
\caption{ In-medium contribution to the width of the $\omega$ meson at zero momentum 
due to its coupling to $K\bar{K}$ (left plot) and the s-wave $\omega N \to VB$ interaction (right plot),  at $\rho_0$ and as a function of the $\omega$ energy $P^0$. 
}
\label{fig:4}
\end{center}
\end{figure}

In left plot of Fig.~\ref{fig:4} we show the in-medium $\omega$ width correction coming from its coupling to
$K\bar{K}$ states in matter. At normal nuclear saturation density, $\rho_0=0.17 {\rm fm}^{-3}$, and around the free $\omega$ mass,  this amounts for 2.9 MeV for an $\omega$ meson at rest. This correction to the width mainly comes from the
$\omega N -K Y$ processes, with $Y=\Lambda (\Sigma)$, that result from the $p$-wave coupling of $\bar K$ to
$YN^{-1}$.

We also present in the right plot of Fig.~\ref{fig:4} the $\omega$ width correction associated
to the elastic and inelastic processes from the $s$-wave interaction of
$\omega N$ with vector mesons and baryons as a function of
the $\omega$ energy. We observe that this contribution
produces a very small
$\omega$ width correction, about $0.5$~MeV,
for energies around the free
$\omega$ mass and at $\rho_0$. The small $\omega$ width correction is associated to the  $\omega N \to \omega N$ and $\omega N \to  \rho N$ processes. Note that the 
implementation of pseudoscalar mesons, hence opening vector-baryon to pseudoscalar-baryon transitions such as $\omega N \to \pi N$, might also add some width to the $\omega$ decay in matter. For that purpose, we adopt the model independent view of  Ref.~\cite{Friman:1997ce},  based on detailed balance and unitarity, and add 9 MeV to the width of the $\omega$ meson.

\begin{figure}[t]
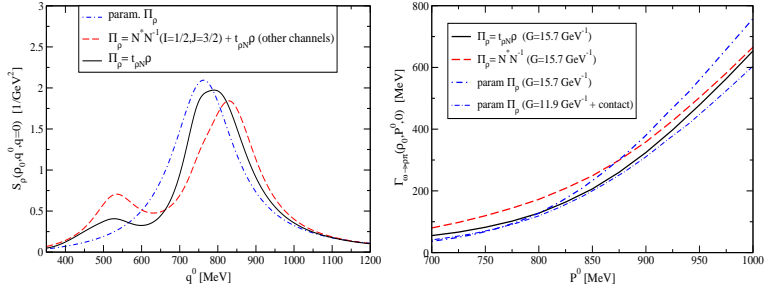

\begin{center}
\includegraphics[width=0.33\textwidth,height=0.25 \textwidth]{fig3_rho_prop_paper.eps}
\includegraphics[width=0.33\textwidth,height=0.25 \textwidth]{fig10_width_energy.eps}
\caption{Left plot: The spectral function of a $\rho$-meson of zero momentum at $\rho_0$ for the three prescriptions employed in this paper. Right plot: Width of the $\omega$ meson from $\omega \rightarrow 3 \pi$ at $\rho_0$ and
$\vec{P}=0$ as a function  $P^0$. 
}%
\label{fig:width_energy}%
\end{center}
\end{figure}

In the right plot of Fig.~\ref{fig:width_energy} we show the in-medium width of an $\omega$ meson at rest for $\rho_0$ as a function of the energy,
from the $\omega \to 3 \pi$ mechanism, which corresponds to absorption processes of the type $\omega N \to \pi \pi N$ and $\omega N N \to \pi N N$. Results are shown for three different prescriptions of the $\rho$ spectral function, displayed on the left of Fig.~\ref{fig:width_energy}, corresponding to using a phenomenological width (dash-dotted line);  employing the $t_{\rho N\to \rho N}$ model from the coupled channel unitary model within the local hidden gauge formalism of Ref.~\cite{Oset:2009vf} but replacing the $I=1/2,J^P=3/2^-$ amplitude by the $N^*(1520)N^{-1}$ contribution of Ref.~\cite{Cabrera:2000dx} (dashed line); and taking the complete  $t_{\rho N\to \rho N}$ amplitude from Ref.~\cite{Oset:2009vf} (solid line). The $\omega \to \rho \pi$ coupling of  $G=15.7$ GeV$^{-1}$ has been adjusted to reproduce the complete free $\omega \to 3 \pi$ width directly from the $\rho \pi$ mechanism. The in-medium $\omega$ width increases smoothly with energy for all the $\rho$-dressing models employed, the phenomenological one (thick dash-dotted line)  presenting a stronger dependence. In this case, results are also shown for the model that uses a contact  term without adjusting the coupling $\omega \to \rho \pi$ of $G=11.9$ GeV$^{-1}$ (thin dash-dotted line). We observe that, up to the free $\omega$ mass, both models present a similar behavior. We conclude that the in-medium width correction at the free $\omega$ mass is $101.2~\pm~10$~MeV for the most complete $\rho$ self-energy model adjusting the $\omega \to \rho \pi$ coupling (solid line), the error associated to reasonable variations in the parameters of the $\pi$ meson self-energy  \cite{omega}.

 In summary, we find  \cite{omega} that the width of the
$\omega$ meson at rest in nuclear matter at saturation density is
$\Gamma_{\omega}(\rho_0,m_\omega)=7.6$~MeV (free width)$ +
101.2$~MeV ($\omega N \to \pi\pi N, \omega N N \to \pi NN$)$+ 2.9$~MeV
($\omega N \to K Y$)$+ 0.5$~MeV ($\omega N \to K^* Y \to \rho N$)$ +9$~MeV
($\omega N \to \pi N$)$= 121~\pm~10$ MeV. We note that one
could add one more MeV to account for the other free decay channels of the
$\omega$ meson,  $\omega \to \pi^0 \gamma$ and $\omega \to \pi^+ \pi^-$.  With regards to the mass shift, no clear conclusion can be drawn due to the uncontrolled high-momentum components of the $\pi$ and $\rho$ propagators \cite{omega}. 

Our value of the width of the $\omega$ meson at rest in nuclear
matter is larger than that found by other works  \cite{Klingl:1998zj,Post:2000rf,Riek:2004kx}, and similar to more recent calculations \cite{Cabrera:2013zga}. In order to compare with the experimental determination of the $\omega$ width, we need to extend our calculation to finite momentum. We find that $\Gamma_{\omega\to 3\pi}$ rises smoothly with momentum, and it can reach values of about 200 MeV at $P=$ 600 MeV/c.  The
experimental width is quoted to be $\Gamma_{\omega} \approx 130-150$~MeV for an
average 3-momentum of 1.1~GeV/c \cite{Trnka:2005ey}. We obtain a good agreement within errors for 400~MeV/c and 600~MeV/c reported in Fig.~4 of Ref.~\cite{Trnka:2005ey},  where our results should be more accurate.

\end{document}